\documentclass[journal]{IEEEtran}

\usepackage[utf8]{inputenc}
\usepackage[T1]{fontenc}
\usepackage[english]{babel}

\usepackage{amsmath,amsfonts}
\usepackage{array}
\usepackage{nicefrac}

\usepackage{multirow, multicol}
\usepackage{enumitem}
\usepackage{comment}

\usepackage{url}

\usepackage{graphicx}
\usepackage{cite}
\usepackage[hidelinks]{hyperref}

\newcommand{\ita}{\textit}
\newcommand{\mrm}{\mathrm}

\newcommand{\mcl}{\mathcal}

\graphicspath{{"Images/"}}

\begin{document}

	\title{Detection of Thermal Events by Semi-Supervised Learning for Tokamak First Wall Safety}
 
        \author{Christian Staron, Hervé Le Borgne, Raphaël Mitteau, Erwan Grelier, Nicolas Allezard

		\thanks{This paper was produced by the Commissariat à l'Énergie Atomique et aux énergies alternatives (CEA -- French Alternative Energies and Atomic Energy Commission), France. We acknowledge the financial support of the Cross-Disciplinary Program on Numerical Simulation of CEA.
		This work was granted access to the HPC resources of IDRIS under the allocation 2022-AD010513390R1 made by GENCI. We also thank Laurent Letellier and Patrick Sayd who were the source of this collaborative work.

  Christian Staron, Raphaël Mitteau and Erwan Grelier are with the IRFM, an institute part of the CEA, F-13108 St Paul Lez Durance, France

  Hervé Le Borgne and Nicolas Allezard are with the List, an institute part of the CEA and Université Paris-Saclay, F-91120, Palaiseau, France.}

\thanks{}}

% The paper headers

\markboth{}
{Staron \MakeLowercase{\textit{et al.}}: Detection of Thermal Events by Semi-Supervised Learning for Tokamak First Wall Safety}

\IEEEpubid{}

\maketitle

\begin{abstract}
This paper explores a semi-supervised object detection approach to detect thermal events on the internal wall of tokamaks. 
A huge amount of data is produced during an experimental campaign by the infrared (IR) viewing systems used to monitor the inner thermal shields during machine operation. The amount of data to be processed and analyzed is such that protecting the first wall is an overwhelming job. Automatizing this job with artificial intelligence (AI) is an attractive solution, but AI requires large labelled datasets that are not readily available for tokamak walls. Semi-supervised learning (SSL) is a possible solution to being able to train deep learning models with a small amount of labelled data and a large amount of unlabelled data. SSL is explored as a possible tool to rapidly adapt a model trained on an experimental campaign \ita{A} of tokamak WEST to a new experimental campaign \ita{B} by using labelled data from campaign \ita{A}, a little labelled data from campaign \ita{B} and a lot of unlabelled data from campaign \ita{B}. Model performance is evaluated on two labelled datasets and two methods including semi-supervised learning. Semi-supervised learning increased the mAP metric by over six percentage points on the first smaller-scale database and over four percentage points on the second larger-scale dataset depending on the method employed.
\end{abstract}

\begin{IEEEkeywords}
Fusion reactors protection, infrared thermography, semi-supervised learning, domain adaptation, object detection
\end{IEEEkeywords}

\section{Introduction}

\IEEEPARstart{T}{he} search for abundant, cheap, low-carbon energy is driving research into nuclear fusion. This reaction, in which two atomic nuclei assemble to form a heavier nucleus while releasing a large amount of energy, is naturally at work in the sun and can be carried out on Earth within tokamaks~\cite{F_Hofmann_1994, y_shimomura_iter_1999, overview_east_2007, Bucalossi_et_al_2022} or stellarators.
Such devices confine a plasma in a limited volume at very high temperatures thanks to powerful magnetic fields. 

\begin{figure}[!t]
\centering
\includegraphics[width=0.9\linewidth]{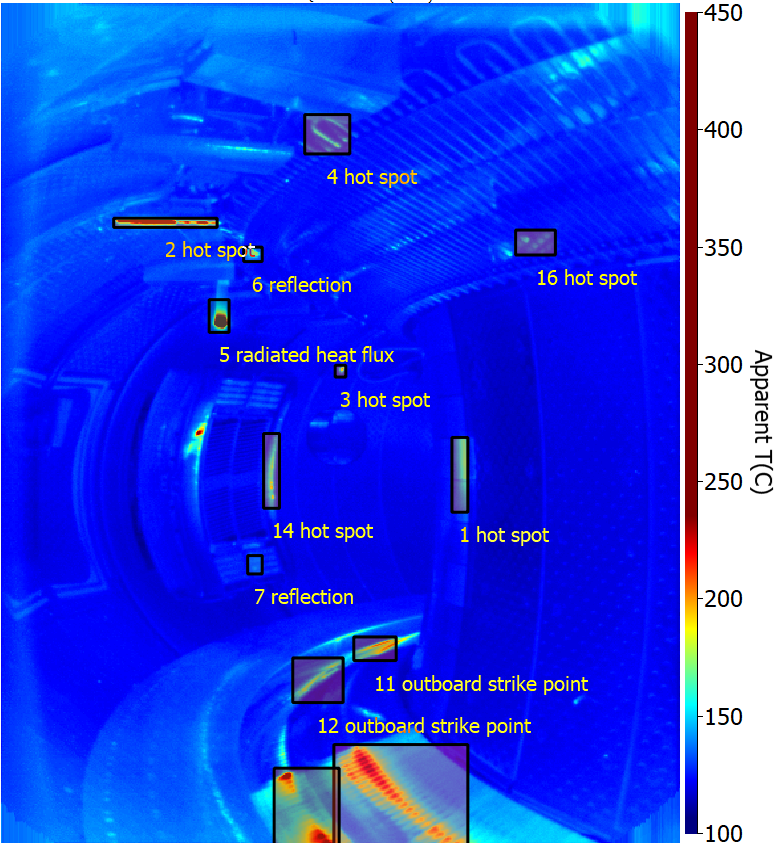}
\caption{Annotation examples of several classes of hot spots on the wide-angle line of sight of the tokamak WEST~\cite{Bucalossi_et_al_2022}}
\label{fig:annotation_examples_wa}
\end{figure}

However, some particles escape inevitably from confinement and reach the Plasma Facing Components (PFC), leading to intense though localized heating. Thermal events appear on the first wall and can damage these in-vessel components when exceeding armour material allowable. Among numerous sensors to monitor the first wall, infrared (IR) viewing systems provide the most relevant information about the surface temperature of the inner walls of tokamaks (\autoref{fig:annotation_examples_wa}). 

The live image feeds are used for real-time feedback control during operation.These data are also used for scientific post-processing after the experiments, to localize and characterize the different classes of thermal events that occur during machine operation. This task requires long and tedious work by high-level experts of plasma physics who would benefit from automation through computer vision approaches. 

First attempts in this direction were based on simple thresholds~\cite{travere2007overview_img_ir_tokamak}  but were tedious to tune manually and suffered from high sensitivity to false detection. It has been slightly improved with the modelling of the background by clustering~\cite{martin2009thermal_event_detect}, but had still limited performance in practice. More recent approaches rely on large neural networks learned from visual data. 
Grelier, Mitteau and Moncada~\cite{grelier_deep_2022} trained a Cascade R-CNN model~\cite{cai201_cascade_rcnn} from scratch using a dataset of 325 thermal events distributed in seven classes, manually annotated from 20 infrared movies from the WEST tokamak. They improved their approach more recently~\cite{Grelier_Mitteau_Moncada_2023} with a Faster R-CNN model~\cite{NIPS2015_14bfa6bb}. Szucs \textit{et al}.~\cite{szucs2022deep_hotspot_W7X} used the YOLOv5~\cite{jocher2020yolov5} model, with a backbone pre-trained on ImageNet and fine-tuning with 471 real images from the W7-X stellarator and 250 synthetic images of hot spots. Such approaches nevertheless still require tedious manual work to annotate the IR video films.

\IEEEpubidadjcol

Using a dedicated tool (described in~\cite{grelier_deep_2022}), labelling one IR film takes between 20 and 30 minutes. Hence, labelling the whole database with these tools manually would take over three years as the database contains more than 13 000 films.

Even by taking multiple precautions, the process of annotation is known to be subject to multiple potential biases~\cite{fabbrizzi2022survey_bias_datasets}. The main potential subject for thermal event labelling is the \ita{label bias} because ``different annotators may assign different labels to the same type of [thermal event]''~\cite{efros2011unbiaised_look}. Since the knowledge of plasma fusion progresses with time, even the same experts can make subtle changes in their assessments to categorize thermal events, between two annotation sessions several months apart.

This article addresses the two aforementioned limitations of the current approaches and proposes an automated process to detect and classify thermal events on the internal wall of tokamaks, which requires a  significantly lesser amount of manual work while maintaining efficiency. We rely on the recent semi-supervised approach SoftTeacher~\cite{xu2021end} to take advantage of the large amount of unlabelled data produced by the IR cameras, in the vein of the semi-supervised principles~\cite{ouali_overview_2020,karaliolios2023gpl}. While still requiring a limited quantity of manual annotations, this approach reduces their amount by one to two orders of magnitude, with a controlled and limited drop in performance.

The study also shows a phenomenon that was previously unidentified. From an experimental campaign to the following one, small changes in the tokamak's configuration make the images different without being clearly visible to human experts. However, although difficult to see ``humanly'', these changes cause a collapse of the performance with a supervised approach, learned during an acquisition campaign and applied as is to a subsequent campaign. The method proposed here overcomes this issue with a minimal manual annotation effort on a novel campaign and adapts the model to the (humanly imperceptible) new visual domain.

To summarize, the contributions of this paper are:
\begin{itemize}
    \item a novel method for visual monitoring of the inner wall of tokamaks, based on semi-supervised learning. Through this paradigm, we show it addresses two major limits of the previous fully-supervised approaches, namely the need for massive annotated data and the subjectivity of the human annotator that introduces noise to the labels;
    \item the identification of a novel phenomenon that arises between two acquisition campaigns in a tokamak. While difficult to see, even for human experts, it has a major impact on fully-supervised approaches. We show that our SSL-based approach is much more robust to this issue.
\end{itemize}

In the following, the method based on semi-supervised learning is described in \autoref{sec:method_ssl}. The data handled in the study, which come from two different acquisition campaigns, as well as several experimental results are reported in \autoref{sec:experimental_results}, showing the interest of the semi-supervised approaches with regards to the compromise between manual annotation effort and detection performance on the one hand, and the ability to adapt to a new domain resulting from a new acquisition campaign on the other hand. Finally, we discuss the limitations of the method and possible future impact in \autoref{sec:conclusions}.

\section{Proposed Approach}\label{sec:method_ssl}

Detecting and classifying thermal events on the internal wall of tokamaks is processed here as an object detection problem. Given an input image, an object detection model predicts a collection of bounding boxes that localize and classify the identified hot spots in the image. Each bounding box consists of 5 predictions: $(x, y, w, h)$ and the confidence for the class label. The $(x, y)$ coordinates represent the upper-left corner of the box, $w$ its width and $h$ its height.

Most recent work proposed to estimate such a model in a supervised learning setting~\cite{grelier_deep_2022,szucs2022deep_hotspot_W7X,Grelier_Mitteau_Moncada_2023}. They rely on a large labelled dataset $L^\mrm{sup}_\mrm{train}$ and minimize a loss, whose exact form depends on the outputs design of the model. In that case, the loss is the sum of a cross-entropy loss for classification and a smooth $L_1$ loss for regression. The regression allows us to estimate the values $(x,y,w,h)$ for each bounding box, while the classification loss results in an estimation of the confidence over the class label. The final loss $\mcl{L}^\mrm{sup}$ is a weighted sum of all these losses, that is minimized with stochastic gradient descent. % (SGD).

\subsection{Semi-supervised learning to detect thermal events}

The main limit of the supervised approach is that it requires a large dataset $L_{\mrm{train}}^{\mrm{sup}}$, which is tedious to manually annotate. The Semi-Supervised Learning (SSL) paradigm is adopted as it needs a much smaller labelled training set $L_{\mrm{train}}$ but takes advantage of a large unlabelled training set $U_{\mrm{train}}$ that is obtained at marginal cost during each acquisition campaign of plasma fusion within the tokamak.
The general process of SSL on infrared images encompasses two steps:
\begin{enumerate}[label=\arabic*.)]
\item Step 1:  (burn-in phase) Train the model on annotated images belonging to $L_{\mrm{train}}$. The burn-in phase is a fully supervised learning phase, also used as a reference later in the paper
\item Step 2:  %(semi-supervised phase)
Apply the model to the images of $U_{\mrm{train}}$ to estimate a pseudo-label for each of them. Then train the model with both the labelled images from 
$L_{\mrm{train}}$ and the pseudo labelled images from $U_{\mrm{train}}$
\end{enumerate}

Since the size of $L_\mrm{train}$ is typically 1\% of that of $L_\mrm{train}^\mrm{sup}$, it reduces significantly the manual effort of annotation required. On the other hand, the size of $U_\mrm{train}$ is of the same order as that of $L_\mrm{train}^\mrm{sup}$ but since the pseudo-labels are obtained automatically, it does not require manual effort.

In the vein of the recent literature in SSL for object detection~\cite{sohn2020simple, Liu_et_al_2021, xu2021end} this work relies on a student-teacher architecture that comprises two branches. The teacher model receives the input of unlabelled images weakly augmented and annotates them with pseudo labels. On the other branch, the input images are strongly augmented and the student model uses these images and the estimated pseudo labels to train the hot spot detector itself. During training, the student model is trained by minimizing the weighted sum of a supervised loss $\mathcal{L}^\mrm{sup}$ estimated with labelled images and an unsupervised loss $\mathcal{L}^\mrm{unsup}$ that results from unlabelled images only. In practice, $\mathcal{L}=\mathcal{L}^\mrm{sup}+\lambda\mathcal{L}^\mrm{unsup}$, where $\lambda$ is a hyperparameter that is 0 during the burn-in phase and is non-null during Step 2. The teacher model is an exponential mean average (EMA)~\cite{Tarvainen_Valpola_2018} of the student model. Its parameters $\theta^\mrm{teacher}$ are updated according to $\theta^\mrm{teacher}\leftarrow \alpha\theta^\mrm{teacher}+(1-\alpha)\theta^\mrm{student}$  at each iteration, where $\alpha$ is a hyperparameter slightly less than one
Note that to detect an object in an image, we face a class imbalance between the boxes that contain a potential object (foreground) and those that do not (background). This imbalance is managed by object detectors such as Faster-RCNN,  which is used at the core of our approach, thanks to a binary detector of \textit{objectness}, that acts on the last convolutional layer of the backbone, which is further used to detect (through regression) and recognize the class of the hot spot (through classification).

Applying data augmentations to input images is a well-known efficient strategy in SSL\cite{Xie_et_al_2020}. They enable a diversification particularly of the unlabelled data so that the teacher model does not fall into the trap of always predicting the same boxes at each iteration. Different augmentations are applied depending on the set from which the data is from.

\begin{itemize}
\item Weak augmentations are applied to unlabelled images that are given as input to the teacher model. Weak augmentations include resizing the image or flipping it in relation to the vertical axis that separates the image in half.
\item Supervised augmentations are applied to the labelled images given as input to the student model. Supervised augmentations add to weak augmentations one extra augmentation that is picked from a list of augmentations with RandAugment \cite{cubuk2019randaugment}. The image can be for instance sharpened, equalized, brightened, contrasted or even not augmented at all.
\item Strong augmentations are applied to the unlabelled images given as input to the student model. Strong augmentations add to the already applied supervised augmentations transformations like translations, shearing, rotating or erasing parts of the image by Cutout \cite{devries2017improved}.
\end{itemize}

A choice was made on which augmentations could be applied to WEST's infrared images because the pixel values in the IR images have a physical interpretation, namely the surface temperature measurements. All augmentations listed in \cite{xu2021end} that strongly affect the pixel values are thus not applied. This includes augmentations that equalize the image, solarize it or mix the color channels. The weak augmentations are unchanged. They are the same augmentations used when training a supervised model in Step 1. The kept supervised augmentations comprise of augmentations that contrast the images, brighten or sharpen them. Strong augmentations are linked to more physical augmentations. The preserved augmentations consist of translating, shearing and erasing part of the images at each iteration.

\subsection{Adaptation to a novel acquisition campaign}\label{sec:ssl_for_campaign_adaptation}

As mentioned above, there may be a difference between images generated by IR cameras for subsequent acquisition campaigns. These differences are either imperceptible to the human eye or blinded by the human brain which tends to correct the images unconsciously. Differences can be due to the installation of new components or various plasma configurations that affect the location of the power deposition. 

In this paper, the images are collected from two such campaigns, denoted by \ita{A} and \ita{B}, described in \autoref{sec:datasets}. As shown experimentally in Section~\ref{sec:model_perf_between_campaigns_and_datasets}, when a detection/classification model is trained with data from campaign \textit{A}, its performance drops dramatically when applied directly on data from campaign \textit{B}. For a new campaign, the parameters $\theta_A$ previously learned need to be adapted to new parameters $\theta_B$ that provide acceptable performance. Such an issue can be seen as a domain adaptation problem~\cite{farahani2020domain_adapt_review} that can be addressed with transfer learning approaches such as using universal backbones~\cite{tamaazousti2017mucale, tamaazousti2019universal}.

Fine-tuning is a common method that enables the quick adaptation of model weights. Hence, one can therefore adjust a model's weights when trained on campaign \ita{A} with images from campaign \ita{B} to quickly gain good performance on a test set containing exclusively images from campaign \ita{B}. However, it still requires a significant amount of labelled data to avoid overfitting and reach the performance of a model that would be learned from scratch on images from both \textit{A} and \textit{B}.

This article aims at demonstrating that an approach based on SSL is more efficient than fine-tuning for a given budget of labelled images. The two following scenarios are considered in the following:

\begin{itemize}
    \item method 1: \label{sth:method1} the initial burn-in is performed from scratch with a small amount of labelled images, either from campaign \textit{A} or \textit{B}. Then, Step 2 is performed with these images and unlabelled images from \textit{B} only.
    \item method 2: \label{sth:method2} the initial burn-in is performed from scratch with a small amount of labelled images from campaign \textit{A} then fine-tuned with images from \textit{B}. Then Step 2 is performed similarly to method 1, with these images and unlabelled images from \textit{B} only.
\end{itemize}

Method 1 reflects an approach that requires retraining the model at each new campaign, with a small amount of labelled data. Method 2 is closer to an ideal realistic use case for WEST's operating team, in which the model learned previously in campaign \textit{A} is directly adapted to campaign \textit{B} without training from scratch.

\subsection{Implementation details}

\begin{figure}[!t]
\centering
\includegraphics[width=0.9\linewidth]{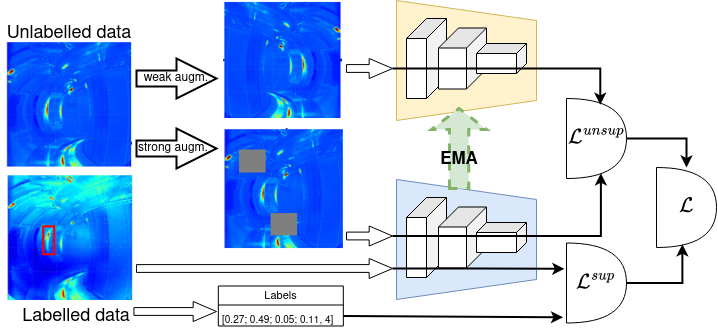}
\caption{The architecture of the model used in this study. The student model (bottom, in blue) learns a supervised loss $\mathcal{L}^{sup}$ with labelled data and an unsupervised loss $\mathcal{L}^{unsup}$ with unlabelled data, which the pseudo-label is provided by the teacher (top, yellow). At each epoch, the teacher is updated from the weights of the student with an exponential moving average (EMA). More details can be found in~\cite{xu2021end} which code was used at the base of our implementation}
\label{fig:hodetec_archi}
\end{figure}

The implementation of our model (\autoref{fig:hodetec_archi}) is built upon the code of SoftTeacher~\cite{xu2021end}, that itself relies on MMDetection~\cite{chen_mmdetection:_2019}. The (student) detection model is a Faster R-CNN with a ResNet-50 \cite{He_Zhang_Ren_Sun_2016} backbone. Such a model is thus similar to that used by recent works in a supervised setting~\cite{Grelier_Mitteau_Moncada_2023}. Hence, the supervised loss $\mathcal{L}^{sup}$ is the sum of a cross entropy loss for classification and a smooth $L_1$ loss for regression, computed with labelled data only. The unsupervised loss $\mathcal{L}^{unsup}$ is also the sum of a classification and a regression losses, but the annotation is provided through a pseudo-label computed by the soft-teacher. The unsupervised classification loss is modified such that the background boxes are weighted according to their reliability score, while the unsupervised regression loss is modified to filter the 2D boxes that the teacher model does not regress with enough stability after a small jittering~\cite{xu2021end}. \autoref{tab:hyperparameters} shows the main hyperparameters used to train both the supervised models and the semi-supervised models. 
The setup of the hyperparameters results either from the values proposed by \cite{xu2021end} or thanks to a grid search conducted on a \nicefrac{2}{3} / \nicefrac{1}{3} split of the training dataset of $\mathcal{D}^1$ (described in \autoref{sec:datasets}). Hence, the relative weight $\lambda$ of the unsupervised loss is fixed at $4.0$ during Step 2.
{ The batch sizes are reduced to retain the number of images per GPU while taking into account the number of GPU used. The learning rates are first adapted proportionally to the changes in batch sizes to keep their ratio constant, then further reduced if needed to ensure convergence.}
The models are trained on 4 NVIDIA Tesla V100 16 GB GPU, each computing a quarter of the batch size before it is aggregated to estimate the gradient.

\begin{table}[!t]
\caption{Main hyperparameters adopted to train Faster R-CNN and SoftTeacher models}\label{tab:hyperparameters}
\centering
\renewcommand{\arraystretch}{1.2}
\begin{tabular}{|c|c|c|}
	\hline
	& Supervised training & SSL training \\
    & (burn-in) & \\
	\hline
   Relative weight $\lambda$ &  -- & 4.0 \\
    \hline
	Learning rate & $5\times 10^{-3}$ & $5\times 10^{-4}$ \\
	\hline
	Batch size    & 16                & 20 \\
	\hline
	Total iterations & \multicolumn{2}{c|}{80 000} \\
	\hline
	Scheduler & \multicolumn{2}{c|}{40 000 and 53 333 iterations} \\
	\hline
	Weight Decay & \multicolumn{2}{c|}{0.0001} \\
	\hline
	Momentum & \multicolumn{2}{c|}{0.9} \\
	\hline
        EMA update $\alpha$ & \multicolumn{2}{c|}{0.999} \\
    \hline
\end{tabular}
\end{table}

The supervised models of the burn-in stage are trained on a number of images from campaign \ita{A} equivalent to 1\% of the fully labelled dataset and a number of images from campaign \ita{B} corresponding to 0.5\% of the labelled dataset. Models are often trained on 10\%, 5\% and 1\% of labelled data in literature \cite{sohn2020simple, Liu_et_al_2021}, usually on the RGB images of MSCOCO \cite{lin2014mscoco}. The decision to train models on 1\% relies on the fact that it is the hardest case because the least amount of labelled data is dealt with during training but also it is the most realistic use case for WEST, as a lot of unlabelled data is at disposal.

\section{Experimental results on WEST's datasets}\label{sec:experimental_results}

\subsection{WEST Dataset}\label{sec:datasets}

The work of this article is based on images from the tangential line of sight of the WEST tokamak \cite{courtois_full_2019,Bucalossi_et_al_2022}. The line is  equipped with an infrared camera, and provides an overall wide-angle (WA) view of the WEST tokamak. It has thus a key role in the protection of the machine because it has the largest variety of thermal events (\autoref{fig:annotation_examples_wa}).

A WEST fusion experiment lasts typically between 15 and 60 seconds, and the longest one up to several minutes. Acquisition campaigns are organized on a regular basis, separated by several months to service the tokamak and its utilities. The images resulting from the videos are of size $512\times 640$px and encoded over 16 bits.

\begin{figure}[!t]
\centering

\includegraphics[width=0.9\linewidth]{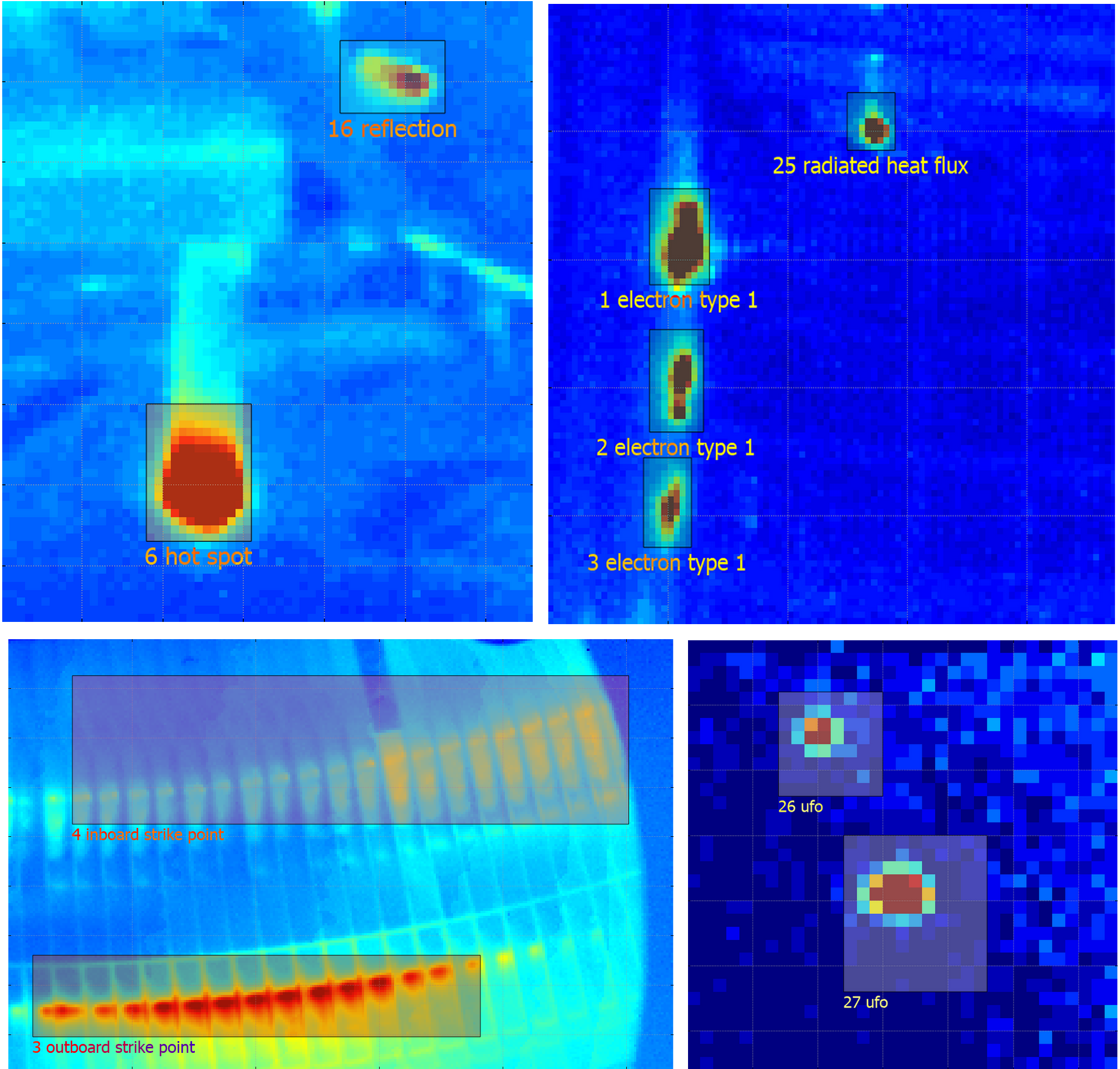}
\caption{Examples of different manual annotations (upper left: reflection and hot spot; upper right: radiated heat flux and electrons type 1; lower left: inboard and outboard strike point; lower right: UFOs)}
\label{fig:different_classes_thermal_events_in_database}
\end{figure}

The thermal events are labelled according to seven classes for the purpose of this work (reported in \autoref{tab:classes_in_dataset}, illustrated in \autoref{fig:different_classes_thermal_events_in_database}, and presented in~\cite{grelier_deep_2022}), while a full data processing pipeline would have many more classes. 
Distinguishing the different categories and assigning a label to a bounding box is a difficult task, that requires a good knowledge of the machine and the use of both the temporal behavior of the thermal event and data from other diagnostics.
Each label consists of four integers corresponding to the surrounding box of the spot, and one to its class. The six first classes correspond to well-defined wall events and the seventh is the catchall class, which is appropriate when the exact definition of the thermal event is undefined. In practice, with progress in plasma fusion concerning the identification of these thermal events, new classes can be defined in the future and include a part of the annotations in this catchall class. Note that the distinction is made between ``thermal events'' and ``hot spots'': a hot spot is related to one single timestamp. By contrast, a thermal event is a sequence of hot spots. This distinction is important because some classes of thermal events can only be determined using spatio-temporal information. For instance, UFOs can be classified as such only by looking at the movement between subsequent hot spots.

\begin{table}[!t]
\caption{The different thermal event classes in the labelled datasets}\label{tab:classes_in_dataset}
\centering
\renewcommand{\arraystretch}{1.2}
\begin{tabular}{|c|c|c|c|}
	\hline
	Class & Class & \multicolumn{2}{c|}{Number of hot spots per class} \\
	\cline{3 - 4}
	number & name & $\mcl{D}^1$ & $\mcl{D}^2$ \\
	\hline
	1 & electron type 1 & 2 103 & 0 \\
	\hline
	2 & inboard strike point & 13 078 & 26 132 \\
	\hline
	3 & outboard strike point & 19 569 & 32 434  \\
	\hline
	4 & reflection & 856 & 724 \\
	\hline
	5 & radiated heat flux & 631 & 0 \\
	\hline
	6 & UFO & 30 & 149 \\
	\hline
	7 & hot spot & 61 237 & 119 751 \\
	\hline	
\end{tabular}
\end{table}

\textbf{Acquisition Campaigns. }
WEST's experimental campaigns last over several months.
WEST goes through many different plasma configurations during these campaigns and evolves from one campaign to another (adding new components, new diagnostics...). 

\begin{figure}[!t]
\centering
\includegraphics[width=\linewidth]{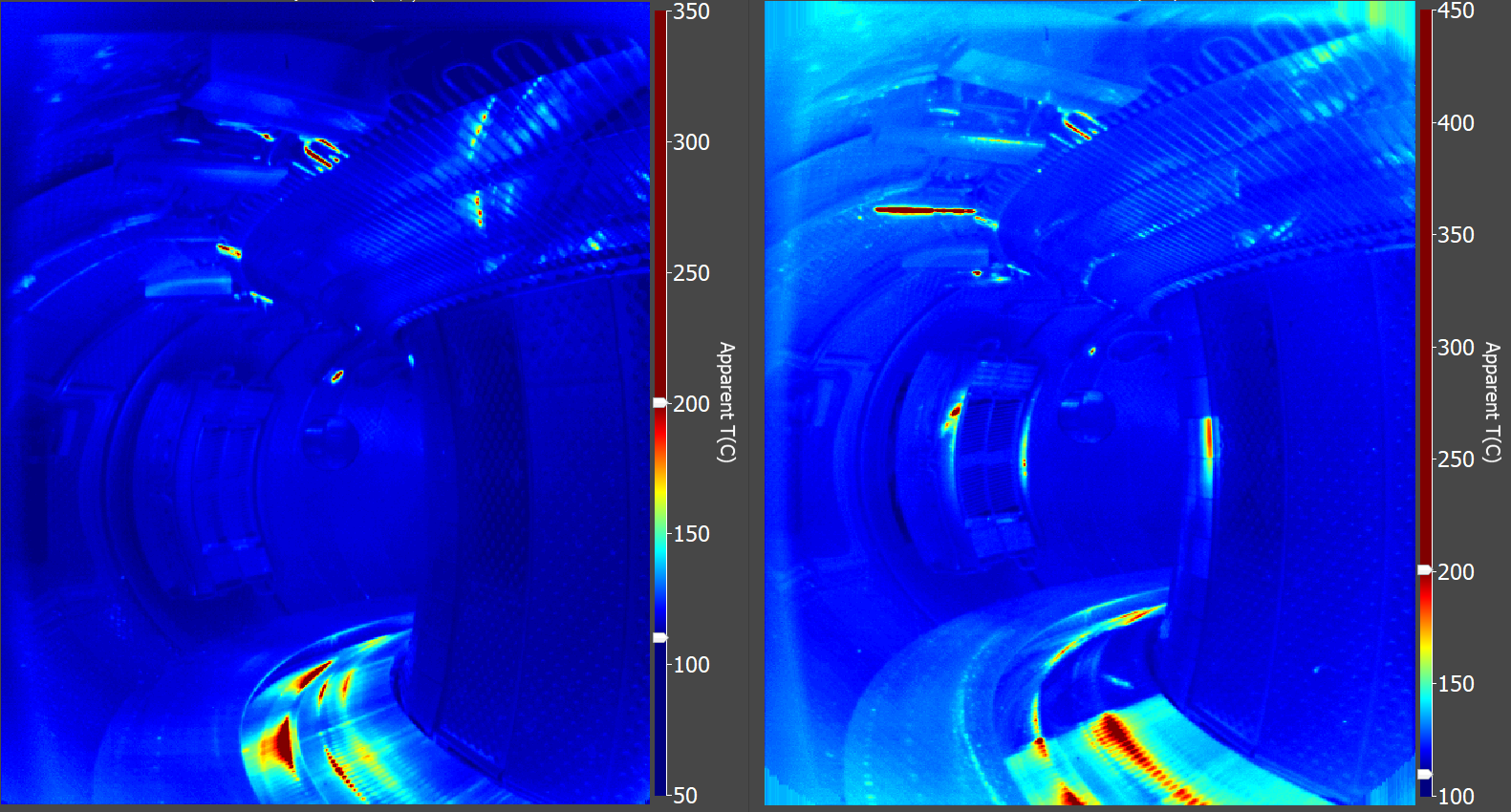}
\caption{Image from campaign \ita{A} (left) during an experiment with similar experimental conditions as the image from campaign \ita{B} (right)}
\label{fig:comparison_images_CA_CB}
\end{figure}

Such changes make the images acquired by the IR cameras different from one campaign to another, as illustrated in \autoref{fig:comparison_images_CA_CB}. Some changes may not be detected by human experts because human tend to correct the images unconsciously, but those changes have a significant impact on the low-level statistics of the images and, as a result, on the performance of the automatic vision systems.

The images we consider were extracted from video films acquired during two different acquisition campaigns to be able to study this effect, namely the campaign \ita{A} which took place between July and November 2019, and the campaign \ita{B} (November 2020 -- January 2021).

\textbf{Annotation Campaigns.}
Two annotation sessions were organized to study potential differences between annotation campaigns. $\mcl{D}^1$ is made of eight video films annotated in September 2021, corresponding to 12 156 images~\cite{grelier_deep_2022}. $\mcl{D}^2$ was annotated by the same expert in August 2022, on 21 other video films, corresponding to 18 336 images~\cite{Grelier_Mitteau_Moncada_2023}. Note that both $\mcl{D}^1$ and $\mcl{D}^2$ contain images from acquisition campaigns A and B. A unique expert annotated both $\mcl{D}^1$ and $\mcl{D}^2$ to mitigate the risk of inter-annotator variation.
The experiments of $\mcl{D}^1$ are picked by experts because they contain interesting events from a machine protection standpoint, whereas the ones of $\mcl{D}^2$ are selected algorithmically to maximize the diversity in machine configurations (plasma configuration, duration, injected power and energy).
\autoref{tab:classes_in_dataset} shows the number of occurrences of each class of hot spot for both datasets. A value of 0 indicates that the dataset does not contain any image that has a given class of thermal event.

\textbf{Splitting for Experiments. } 

Each of the two annotated datasets is divided into a train and a test set. The splits are made at the video film level to avoid data leakage. If a detector were trained with images from a given video film, it would be much easier to recognize hot spots from images of the same video film (even if they are not seen during training), due to the strong temporal coherency within each video film. 
\autoref{tab:nb_labelled_images_per_set_per_dataset} gives the number of images in the training and test sets.
In the following, each subset is referred by the notation $X_\beta^\alpha$, where $X\in\{A, B\}$ is the acquisition campaign, $\alpha \in \{1, 2\}$ refers to the annotation session datasets ($\mcl{D}^1$ or $\mcl{D}^2$) and  $\beta\in\{\mrm{train}, \mrm{test}\}$ is the split. $\bar{X}$ denotes the complement of $X$ with respect to the set of labelled data from the campaign $X$. The considered subsets of $X$ contain images that are equally distributed between the video films. 

The same number of images is extracted from each labelled film to have a well-defined representation of each film in the training set. During experiments, if one needs to train on a number of images equivalent to \textit{e.g.} 1\% of the dataset, 96 labelled images are considered for $\mcl{D}^1$ and 144 for the larger dataset $\mcl{D}^2$.
\begin{table*}[!t]

\caption{ Content of the labelled datasets $\mcl{D}^1$ and $\mcl{D}^2$ used in experiments. Image data come from two acquisition campaigns (A and B) and labels from two annotations campaigns (corresponding to the datasets $\mcl{D}^1$ and $\mcl{D}^2$).} \label{tab:nb_labelled_images_per_set_per_dataset}
\renewcommand{\arraystretch}{1.1}
\centering
\begin{tabular}{|c|c|c|c|c|}
	\hline
	& \multicolumn{2}{c|}{ $\mcl{D}^1$} & \multicolumn{2}{c|}{ $\mcl{D}^2$} \\
	\cline{2-5}
	& Train & Test & Train & Test \\
	\hline
	Number of films from \ita{A} & 2 & 1 & 8 & 1 \\
	Number of films from \ita{B} & 3 & 2 & 9 & 3 \\
	\hline
	Number of images & 7 879 & 4 277 & 15 130 & 3 206 \\
	\hline
	Notation & $A_\mrm{train}^1+B_\mrm{train}^1$ & $A_\mrm{test}^1+B_\mrm{test}^1$ & $A_\mrm{train}^2+B_\mrm{train}^2$ & $A_\mrm{test}^2+B_\mrm{test}^2$ \\
	\hline
\end{tabular}
\end{table*}

\subsection{Evaluation metric}

The mean Average Precision (mAP)
is used to measure the performance. It is a reference metric in object detection that both reflects the localization and classification ability of the model, and aggregates the performance for all the classes considered. The mAP values reported for SSL experiments in the following are always those of the student model. The mAP implementation is the same as the one adopted for the COCO challenges, usually noted $\mrm{mAP}_{0.5:0.95}$, that is the mAP averaged over an ``intersection over union'' (IoU) between the detected hot spot and the ground truth at $0.5, 0.55, 0.6,\dots,0.95$. Following \cite{Zhou_et_al_2021}, the mAP is only computed for boxes with a confidence higher than $10^{-3}$. % Christian: J'ai trouvé cela sur la page Github de SoftTeacher : c'est dans les notes (https://github.com/microsoft/SoftTeacher#notes)

During the preliminary tests of our work, we considered all the standardized COCO metrics, but did not notice any remarkable results both globally or by class.  In other words, the relative order of the performances was always the same regardless of the metric. For this reason, we only report the (integrated) mAP as described above.

\subsection{Inter-campaigns and Annotations Domain Shift}\label{sec:model_perf_between_campaigns_and_datasets}
\hyperref[sec:datasets]{Subsection \ref*{sec:datasets}} explains that the datasets have been created to study the transfer of a model learned on the data of campaign \ita{A} to the test data of a campaign \ita{B}. Firstly, a series of experiments in a fully labelled setting is conducted. The reporting of the results is in \autoref{tab:performance_between_campaigns_and_datasets}.

The results are quite good (lines (i), (iv) and (v) of \autoref{tab:performance_between_campaigns_and_datasets}) when the model is trained and tested on data from the same campaign (and annotated homogeneously in $\mcl{D}^1$: see below). The results on \ita{B}$^1$ are better than on \ita{A}$^1$ but the inclusion of data from \ita{B} boosts performance. This may be due to a larger training dataset or more relevant images in \ita{B}$^1$. However, the performance drops dramatically to near zero (lines (ii) and (iii) of \autoref{tab:performance_between_campaigns_and_datasets}) when the training and test data are from different campaigns. These experiments show that the changes in tokamak configurations from one campaign to another produce subtle changes in the low-level statistics of the images that are not visible to experts (see \autoref{fig:comparison_images_CA_CB}) but have a significant impact on performance, comparable to that of a domain shift.

The datasets enable the study of the influence of a potential light change in the annotation process. Indeed, while both datasets $\mcl{D}^1$ and $\mcl{D}^2$ contain data from campaigns \ita{A} and \ita{B}, $\mcl{D}^2$ is annotated almost one year after $\mcl{D}^1$. Even though the annotators are the same, the criterion to classify the thermal events by experts changed slightly, introducing minor differences between the two.
For instance, following discussions with experts, for $\mcl{D}^2$, the reciprocating Langmuir probe saw its classification change from ``radiated heat flux'' to ``hot spot'', and some thermal events on the upper divertor were annotated and classified as ``reflection''. Furthermore, the differences in diversity of the configuration of the machine, described in \autoref{sec:datasets}, also introduce differences between the two datasets.

Good performance is still attained when the training and test data is homogeneous (lines (vii), (x) and (xi) of \autoref{tab:performance_between_campaigns_and_datasets}) but a significant drop is noted when the training set and the test one come from different annotation sessions (lines (vi) and (viii) of \autoref{tab:performance_between_campaigns_and_datasets}).
The fact that the model trained on $\mcl{D}^2$ has better performance than that trained on $\mcl{D}^1$ shows that later data encapsulates earlier ones.
The domain shift identified in $\mcl{D}^1$ is nevertheless still present in $\mcl{D}^2$ (lines (ix) to (xii)  of \autoref{tab:performance_between_campaigns_and_datasets}).

\begin{table}[!t]
	\caption{\textnormal{m}AP for different campaigns and datasets}\label{tab:performance_between_campaigns_and_datasets}
	\centering
	\renewcommand{\arraystretch}{1.25}
	\begin{tabular}{|c |c c | c c |}
		\hline
		& \multicolumn{2}{c|}{Initial training}    & \multicolumn{2}{c|}{ Evaluation} \\
		\hline
		Experiment & Number  & \multirow{2}{*}{set} 		    & Test & \multirow{2}{*}{mAP}   \\
		number	   & of images &                              & set & \\
		\hline
        (i) & 5 575 & $A_\mrm{train}^1$ & $A_\mrm{test}^1$ & 0.381 \\
        (ii) & 5 575 & $A_\mrm{train}^1$ & $B_\mrm{test}^1$ & 0.030 \\
        (iii) & 2 304 & $B_\mrm{train}^1$ & $A_\mrm{test}^1$ & 0.037 \\
        (iv) & 2 304 & $B_\mrm{train}^1$ & $B_\mrm{test}^1$ & 0.435 \\
        \hline
		\multirow{2}{*}{(v)} & \multirow{2}{*}{7 879} & $A_\mrm{train}^1$ & $A_\mrm{test}^1$ & \multirow{2}{*}{0.434} \\
		& & $+ B_\mrm{train}^1$ & $+ B_\mrm{test}^1$ & \\
        \hline
		\multirow{2}{*}{(vi)} & \multirow{2}{*}{7 879} & $A_\mrm{train}^1$ & $A_\mrm{test}^2$ & \multirow{2}{*}{0.087} \\ %(ff) was (i)
		& & $+ B_\mrm{train}^1$ & $+ B_\mrm{test}^2$ & \\
		\hline
        \multirow{2}{*}{(vii)} & \multirow{2}{*}{15 130} & $A_\mrm{train}^2$ & $A_\mrm{test}^2$ & \multirow{2}{*}{0.385} \\
		& & $+ B_\mrm{train}^2$ & $+ B_\mrm{test}^2$ & \\
		\multirow{2}{*}{(viii)} & \multirow{2}{*}{15 130} & $A_\mrm{train}^2$ & $A_\mrm{test}^1$ & \multirow{2}{*}{0.168} \\
		& & $+ B_\mrm{train}^2$ & $+ B_\mrm{test}^1$ & \\
		\hline

		(ix) & 4 606 &  $B_\mrm{train}^2$ & $A_\mrm{test}^2$ & 0.030 \\
		(x) & 4 606 &  $B_\mrm{train}^2$ & $B_\mrm{test}^2$ & 0.330 \\

        (xi) & 10 524 & $A_\mrm{train}^2$ & $A_\mrm{test}^2$ & 0.396 \\
        (xii) & 10 524 & $A_\mrm{train}^2$ & $B_\mrm{test}^2$ & 0.017 \\
        \hline

	\end{tabular}
\end{table}

The change in model performance between datasets and campaigns demonstrates that there is a need to quickly adjust model weights when faced with a new experimental campaign. The following subsections explicit how semi-supervised learning can address this need.

\subsection{SSL Gets Near Supervised Performance at Lower Cost}\label{sec:compare_supervised}

This section compares the approach based on semi-supervised learning and the classical supervised setting~\cite{grelier_deep_2022,Grelier_Mitteau_Moncada_2023}. Experiments are conducted both on $\mcl{D}^1$ and $\mcl{D}^2$. The results are reported in respectively \autoref{tab:eal_WA} and \autoref{tab:eal_WA2} with the same experiments for both datasets, numbered from (i) to (vii). The supervised settings is reproduced with the same code as our approach for a fair comparison.

The best possible performance is achieved with the Faster R-CNN model in a supervised setting with a ResNet-50 backbone, when the model is trained with all data (from campaigns \ita{A} and \ita{B}). This performance varies from 0.32 to 0.477 (line (i) in both tables) depending on the test dataset. The performance logically decreases between 0.267 and 0.435 when only a part of the training data is used (line (ii) to (iv)), depending on the exact settings either at training or testing.

The performance drops with regard to the best fully supervised model with the proposed approach based on SSL (line (vi) of both tables). The drop is at most 4 mAP percentage points on $\mcl{D}^1$ with $\frac{7879}{96+48}\approx54$ times less data and 3.7 percentage points on $\mcl{D}^2$ with $\frac{15130}{144+72}\approx70$ times less data. If one uses the same number of annotated data (line (iii) versus (vi)) the SSL approach surpasses the supervised approach. On $\mcl{D}^1$, the SSL approach even surpasses the model trained on all the data from $B_\mrm{train}^1$ (0.456 vs 0.435) when both are tested on $B_\mrm{test}^1$ (line (vi) versus (ii)), although it gets lower ones (0.303 vs 0.33) on $\mcl{D}^2$.
Semi-supervised learning is therefore a way to get near fully supervised performance at a much smaller labelling cost. 

\subsection{SSL Better Adapts than Fine-Tuning}\label{sec:ssl_beats_fine_tuning}

The SSL approach is compared to the classical fine-tuning as a method to adapt the model to a novel campaign. On line (v) of  \autoref{tab:eal_WA} and \autoref{tab:eal_WA2}, a model is first trained during 80 000 iterations with 1\% of data from campaign \ita{A}(96 images for $\mcl{D}^1$, 144 for $\mcl{D}^2$), then fine-tuned during 10 000 iterations with 0.5\% data from campaign \ita{B} (48 images for $\mcl{D}^1$, 72 for $\mcl{D}^2$). Fine-tuning allows the expected adaptation, since the performance on $\ita{B}_\mrm{test}$ is no longer almost null (line (v) versus (iv) of \autoref{tab:eal_WA} and \autoref{tab:eal_WA2}, respectively). There is an expected decrease of performance on $A_\mrm{test}$ as fine-tuning specializes the model's parameters to the new campaign \ita{B} (line (v) of \autoref{tab:eal_WA} and \autoref{tab:eal_WA2}). Fine-tuning remains however inferior to the model trained with the same amount of annotated data from \ita{A} and \ita{B} (line (iii) versus (v) of \autoref{tab:eal_WA} and \autoref{tab:eal_WA2}, respectively).

\begin{figure}
    \centering
    \includegraphics[width=\linewidth]{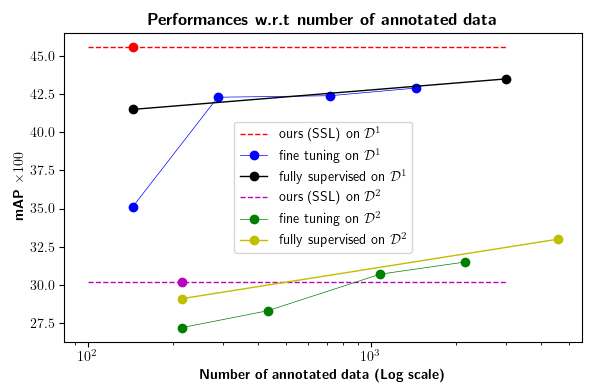}
    \caption{Performance according to the number of annotated data (Log  scale in $x$) for our approach (SSL) compared to fine-tuning and supervised learning, both on $\mcl{D}^1$ and $\mcl{D}^2$. The performance of our approach is given for a unique number of annotated data (points). The dashed line is merely drawn as an aid for figure readability : the dashed line recalls the reference level for the continuous lines}
    \label{fig:ft_vs_ssl}
\end{figure}

For both datasets, the same amount of labelled data is handled but additional unlabeled data is used for the SSL model training. The SSL approach surpasses fine-tuning by 3 to 10 percentage points (line (vi) versus (v)  of \autoref{tab:eal_WA} and \autoref{tab:eal_WA2}, respectively). With a small amount of labelled data representing less than 2\% of all available data\footnote{ 144 annotated images from $\mcl{D}^1$ used in line (vi) of \autoref{tab:eal_WA}, from a total of 7879 images; 216 from $\mcl{D}^2$ used in line (iii) of \autoref{tab:eal_WA2}, from a total of 15130.}, SSL even surpasses the supervised approach on the later campaign $\ita{B}_\mrm{test}$ (line (vi) versus (iii) of \autoref{tab:eal_WA} and \autoref{tab:eal_WA2}, respectively), in accordance with the results of the previous section.
A study including an increased number of annotated images used for fine-tuning to a factor of 2, 5, and 10 times the number of annotated data used with SSL is carried out. The results are reported in \autoref{fig:ft_vs_ssl}. The fine-tuning on $\mcl{D}^2$ requires 4 times more labelled data to reach the same performance as SSL, while on $\mcl{D}^1$, the performance of fine-tuning is still significantly below, even with 10 times more annotated data. We also report the results for the supervised learning approaches with 2304 annotated data on $\mathcal{D}_1$ and 4606 ones on $\mathcal{D}_2$, that are presented in detail in section~\autoref{sec:compare_supervised}.

\begin{table*}[!t]
\caption{Experimental results on $\mcl{D}^1$}\label{tab:eal_WA}
\centering
\renewcommand{\arraystretch}{1.25}
\begin{tabular}{|c |c c | c c | c c |c c|}
	\hline
	& \multicolumn{2}{c|}{Initial training}    & \multicolumn{2}{c|}{Fine-tuning} &  \multicolumn{2}{c|}{Semi-supervised learning}  &     & \\
	\hline
	Experiment & Number  & \multirow{2}{*}{Set} 		    & Number                          & \multirow{2}{*}{Set}                           & Number      & Unlabelled  & Test & \multirow{2}{*}{mAP}   \\
	number	   & of images &                              & of images                       &                                               &  of images    & set      & set & \\
	\hline
	\multirow{3}{*}{(i)} & \multirow{3}{*}{7 879} & $A_\mrm{train}^1$ & \multirow{3}{*}{--} & \multirow{3}{*}{--} & \multirow{3}{*}{--} & \multirow{3}{*}{--} & $A_\mrm{test}^1$ & 0.397 \\
	& & $+ B_\mrm{train}^1$ & & & & & $B_\mrm{test}^1$ & 0.477 \\
	& & & & & & & both & 0.434 \\
	\hline
	(ii) & 2 304 &  $B_\mrm{train}^1$ & -- & -- & -- & -- & $B_\mrm{test}^1$ & 0.435 \\
	\hline
	\multirow{2}{*}{(iii)} & \multirow{2}{*}{96/48} & $A_\mrm{train}^1/$ & \multirow{2}{*}{--} & \multirow{2}{*}{--} & \multirow{2}{*}{--} & \multirow{2}{*}{--} & $A_\mrm{test}^1$ & 0.282 \\
	& & $B_\mrm{train}^1$ & & & & & $B_\mrm{test}^1$ & 0.415 \\
	\hline
	\multirow{2}{*}{(iv)} & \multirow{2}{*}{96} & \multirow{2}{*}{$A_\mrm{train}^1$} & \multirow{2}{*}{--} & \multirow{2}{*}{--} & \multirow{2}{*}{--} & \multirow{2}{*}{--} & $A_\mrm{test}^1$ & 0.267 \\
    & & & & & & & $B_\mrm{test}^1$ & 0.020 \\
	\hline
	\multirow{2}{*}{(v)} & \multirow{2}{*}{96} & \multirow{2}{*}{$A_\mrm{train}^1$} & \multirow{2}{*}{48} & \multirow{2}{*}{$B_\mrm{train}^1$} & \multirow{2}{*}{--} & \multirow{2}{*}{--} & $A_\mrm{test}^1$ & 0.202 \\
    & & & & & & & $B_\mrm{test}^1$ & 0.351 \\
	\hline
	\multirow{2}{*}{(vi)} & \multirow{2}{*}{96/48} & $A_\mrm{train}^1/$ & \multirow{2}{*}{--} & \multirow{2}{*}{--} & \multirow{2}{*}{2 256} & \multirow{2}{*}{$\bar{B}_{\mrm{train}}^1$} & $A_\mrm{test}^1$ & 0.357 \\
	& & $B_\mrm{train}^1$ & & & & & $B_\mrm{test}^1$ & 0.456 \\
	\hline
	\multirow{2}{*}{(vii)} & \multirow{2}{*}{96} & \multirow{2}{*}{$A_\mrm{train}^1$} & \multirow{2}{*}{48} & \multirow{2}{*}{$B_\mrm{train}^1$} & \multirow{2}{*}{2 256} & \multirow{2}{*}{$\bar{B}_{\mrm{train}}^1$} & $A_\mrm{test}^1$ & 0.355 \\
	& & & & & & & $B_\mrm{test}^1$ & 0.416 \\
	\hline
\end{tabular}
\end{table*}

\begin{table*}[!t]
\caption{Experimental results on $\mcl{D}^2$}\label{tab:eal_WA2}
\centering
\renewcommand{\arraystretch}{1.25}
\begin{tabular}{|c|c c | c c | c c |c c|}
	\hline
	& \multicolumn{2}{c|}{Initial training}    & \multicolumn{2}{c|}{Fine-tuning} &  \multicolumn{2}{c|}{Semi-supervised learning}  &     & \\
	\hline
	Experiment & Number  & \multirow{2}{*}{Set} 	  & Number                          & \multirow{2}{*}{Set}                           & Number      & Unlabelled  & Test & \multirow{2}{*}{mAP} \\
	number & of images &                              & of images                       &                                               &  of images    & set      & set &  \\
	\hline
	\multirow{3}{*}{(i)} & \multirow{3}{*}{15 130} & $A_\mrm{train}^2$ & \multirow{3}{*}{--} & \multirow{3}{*}{--} & \multirow{3}{*}{--} & \multirow{3}{*}{--} & $A_\mrm{test}^2$ & 0.425 \\
	& & $+ B_\mrm{train}^2$ & & & & & $B_\mrm{test}^2$ & 0.320 \\
	& &  & & & & & both & 0.385 \\
	\hline
	(ii) & 4 606 &  $B_\mrm{train}^2$ & -- & -- & -- & -- & $B_\mrm{test}^2$ & 0.330 \\
	\hline
	\multirow{2}{*}{(iii)} & \multirow{2}{*}{144/72} & $A_\mrm{train}^2/$ & \multirow{2}{*}{--} & \multirow{2}{*}{--} & \multirow{2}{*}{--} & \multirow{2}{*}{--} & $A_\mrm{test}^2$ & 0.389 \\
	& & $B_\mrm{train}^2$ & & & & & $B_\mrm{test}^2$ & 0.291 \\
	\hline
	\multirow{2}{*}{(iv)} & \multirow{2}{*}{144} & \multirow{2}{*}{$A_\mrm{train}^2$} & \multirow{2}{*}{--} & \multirow{2}{*}{--} & \multirow{2}{*}{--} & \multirow{2}{*}{--} & $A_\mrm{test}^2$ & 0.374 \\
    & & & & & & & $B_\mrm{test}^2$ & 0.015 \\
	\hline
	\multirow{2}{*}{(v)} & \multirow{2}{*}{144} & \multirow{2}{*}{$A_\mrm{train}^2$} & \multirow{2}{*}{72} & \multirow{2}{*}{$B_\mrm{train}^2$} & \multirow{2}{*}{--} & \multirow{2}{*}{--} & $A_\mrm{test}^2$ & 0.12 \\
    & & & & & & & $B_\mrm{test}^2$ & 0.272 \\
	\hline
	\multirow{2}{*}{(vi)} & \multirow{2}{*}{144/72} & $A_\mrm{train}^2/$ & \multirow{2}{*}{--} & \multirow{2}{*}{--} & \multirow{2}{*}{4 534} & \multirow{2}{*}{$\bar{B}_{\mrm{train}}^2$} & $A_\mrm{test}^2$ & 0.388 \\
	& & $B_\mrm{train}^2$ & & & & & $B_\mrm{test}^2$ & 0.302 \\
	\hline
	\multirow{2}{*}{(vii)} &\multirow{2}{*}{144} & \multirow{2}{*}{$A_\mrm{train}^2$} & \multirow{2}{*}{72} & \multirow{2}{*}{$B_\mrm{train}^2$} & \multirow{2}{*}{4 534} & \multirow{2}{*}{$\bar{B}_{\mrm{train}}^2$} & $A_\mrm{test}^2$ & 0.372 \\
	& & & & & & & $B_\mrm{test}^2$ & 0.317 \\
	\hline
\end{tabular}
\end{table*}

These experiments on $\mcl{D}^1$ and $\mcl{D}^2$ show that it is in the tokamak operating team's best interest to apply semi-supervised learning rather than fine-tuning.
It should however be noted that the SSL approach includes a burn-in stage that consists in 80 000 iterations with data from both campaigns, while the fine-tuning approach can directly use the model already available from a previous campaign and compute the adaptation during only 10 000 iterations. The need for the burn-in phase can make the SSL less attractive than fine-tuning for practical usage.

\subsection{SSL Enables Rapid Campaign Adaptation}

The combination of fine-tuning with SSL during the second step of learning (method 2 in Section~\ref{sec:method_ssl}) is applied to address the limits of the previous section. It consists in starting from a model available from a previous campaign (campaign \ita{A}), fine-tuning it on the novel campaign (campaign \ita{B}) then continuing to learn with unlabelled data from the novel campaign. Note that the step of fine-tuning is required since the model from the previous campaign is not reliable enough to create good pseudo-labels. The results of this approach are reported on line (vii) of \autoref{tab:eal_WA} and \autoref{tab:eal_WA2}.

The performance is better (on $\mcl{D}^1$) or on-par (on $\mcl{D}^2$) with a supervised approach with the same amount of labelled data (line (iii) of \autoref{tab:eal_WA} and \autoref{tab:eal_WA2}) but do not need to retrain the model from scratch. In comparison to the SSL approach only (method 1, line (vi) of \autoref{tab:eal_WA} and \autoref{tab:eal_WA2}) the results are globally equivalent, depending on the considered test set. As a consequence, this second version of the approach that relies on SSL still surpasses the fine-tuning only approach (line (v) of \autoref{tab:eal_WA} and \autoref{tab:eal_WA2}).

Applying semi-supervised learning is therefore a labelling cost-effective way to adapt a model trained on a certain experimental campaign to a new one and rapidly get good detection performance on images from the new campaign. The pairing of fine-tuning and semi-supervised learning even achieves the best mAP value after an experiment including semi-supervised learning on $\mcl{D}^2$, the larger and more realistic dataset.

\section{Conclusion and perspectives}\label{sec:conclusions}

This article explored the application of semi-supervised learning to infrared cameras used to monitor the inside of fusion reactors during experimental campaigns. Semi-supervised learning can help quickly adapt a model trained on a certain acquisition campaign to a new one with a lesser data labelling cost. This work establishes that semi-supervised learning outperforms fine-tuning when faced with a new campaign, and that semi-supervised learning can achieve near fully supervised performance but with extensively fewer annotations. A little over six mAP percentage points separate at most the performance of the trained semi-supervised model with the performance of the model trained on all data on the labelled datasets of tokamak WEST but this performance is accomplished with  up to 70 times fewer annotations.

Semi-supervised learning is therefore an attractive tool for future scientific experimental campaigns. This article focuses on the wide-angle viewing line of sight of the WEST tokamak. Experiments including semi-supervised learning could be carried out on other lines of sight available among the 12 cameras installed on WEST as a continuation of this work. Other possibilities include adapting a model trained on an experimental campaign of WEST to other fusion devices or in the near future to the International Thermonuclear Experimental Reactor (ITER) \cite{y_shimomura_iter_1999}.

\section*{Data availability}

The data that supports the findings of this study are available from the corresponding author upon reasonable request, according to the CEA general data-sharing framework, or through formalized collaboration agreement.

\bibliographystyle{ieeetr}
\bibliography{HoDetec_article}

\end{document}